\documentclass[aps,superscriptaddress,prb,a4paper,showpacs]{revtex4}

\usepackage{graphicx,amsmath,amssymb}

\newcommand{\be}{\begin{equation}}
\newcommand{\ee}{\end{equation}}
\newcommand{\bea}{\begin{eqnarray}}
\newcommand{\eea}{\end{eqnarray}}

\newcommand{\s}{\sigma}
\renewcommand{\t}{\tau}

\renewcommand{\b}{\beta}

\newcommand{\pt}{\tilde{p}}
\newcommand{\Trsb}{T_{\rm ag}}     
\newcommand{\FTAP}{F_\text{\sc tap}}
\newcommand{\tap}{{\sc tap}}

\begin{document}

\title{Off-equilibrium confined dynamics in a glassy system with
level-crossing states}

\author{Barbara Capone}
\affiliation{Dipartimento di Fisica, CNR-INFM (UdR Roma I and SMC
center), Universit\`a di Roma ``La Sapienza'', P. Aldo Moro 2, I-00185
Roma, Italy}
\author{Tommaso Castellani}
\affiliation{Dipartimento di Fisica, CNR-INFM (UdR Roma I and SMC
center), Universit\`a di Roma ``La Sapienza'', P. Aldo Moro 2, I-00185
Roma, Italy}
\author{Irene Giardina}
\affiliation{Dipartimento di Fisica, CNR-INFM (UdR Roma I and SMC
center), Universit\`a di Roma ``La Sapienza'', P. Aldo Moro 2, I-00185
Roma, Italy}
\affiliation{ISC-CNR, Via dei Taurini 19, I-00185 Roma, Italy}
\author{Federico Ricci-Tersenghi}
\affiliation{Dipartimento di Fisica, CNR-INFM (UdR Roma I and SMC
center), Universit\`a di Roma ``La Sapienza'', P. Aldo Moro 2, I-00185
Roma, Italy}

\begin{abstract}
We study analytically the dynamics of a generalized $p$-spin model,
starting with a thermalized initial condition. The model presents
birth and death of states, hence the dynamics (even starting at
equilibrium) may go out of equilibrium when the temperature is
varied. We give a full description of this constrained out of
equilibrium behavior and we clarify the connection to the
thermodynamics by computing (sub-dominant) {\sc tap} states,
constrained to the starting equilibrium configuration.
\end{abstract}


\maketitle

Many interesting physical systems live for very long times out of
equilibrium, and, in this regime, they display highly non trivial
behaviors which are still to be understood (e.g.\ rejuvenation and
memory effects in spin glasses).  In general, these systems fall out
of equilibrium when some external parameter is changed. For example,
fragile glass-forming liquids undergo a dramatic slowing down of their
relaxational dynamics when the temperature is dropped below the glass
transition temperature \cite{VetriReview}. This effect is sharpened in
certain mean-field models where, at a critical temperature $T_d$, a
transition occurs from an equilibrium kind of dynamics to an
off-equilibrium aging one \cite{DynamicsReview}.  The phenomenon is
ubiquitous and can be found also in very different fields: e.g.\ in
local search algorithms for solving hard optimization problems the
time-complexity may become extremely large by varying a macroscopic
parameter \cite{CSReview}.  A better understanding of the mechanisms
leading to the dramatic slowing down in out of equilibrium dynamics is
a subject of broad interest and wide applicability.

In describing the dynamical slowing down (and possible eventual
arrest) the common view suggests that at a low temperature a huge
number of metastable states appears (with energies higher than the
equilibrium one, $E_0$), making relaxation to equilibrium very slow,
and even impossible if interactions are long ranged and metastable
states life-times diverge in the thermodynamic limit.  This picture
has been verified by solving the out of equilibrium Langevin dynamics
of a particularly simple mean-field model, the so-called fully
connected spherical $p$-spin, whose Hamiltonian is
\cite{CrisantiSommers_Statics}
\begin{equation}
\mathcal{H}\{\underline\s\} = -\sum_{1 \le i_1 < \ldots < i_p \le N}
  J_{i_1 \ldots i_p} \s_{i_1} \ldots \s_{i_p}\;,
\label{eq:HamSingle}
\end{equation}
where the $N$ spins $\s_i$ are continuous variables subject to the
spherical constraint $\sum_i \s_i^2 = N$ and the couplings are i.i.d.\
random variables with zero mean and variance $p!/(2N^{p-1})$.  In this
model (hereafter $p \ge 3$) if we consider a {\it quench}, that is if
we choose an initial configuration of high energy and let the system
relax at a {\it fixed} value of the temperature $T<T_d$ ($T_d$ being
the dynamic transition temperature), the asymptotic dynamics remains
trapped at the energy level of the highest and most numerous
metastable states, the so-called {\it threshold
states}. Time-translation invariance and the dynamic
fluctuation-dissipation relation are violated and aging is observed in
correlation and response functions \cite{CuKu}.

These features are intriguing and experimentally relevant, since aging
behaviour has been observed in many disordered systems.  Nevertheless,
in order to compare with more realistic situations it is extremely
useful to understand the dynamical behaviour during a {\it cooling}.
In this case the system relaxes while the temperature is slowly
decreased from an initial high value to a final temperature below
$T_d$.

In the single-$p$-spin model, defined by the Hamiltonian
(\ref{eq:HamSingle}), changing the temperature during the dynamics has
no effect on the asymptotic states approached by the relaxing system
\cite{BarratBurioniMezard}.  This can be easily understood by
considering the structure of metastable states.  Thanks to a
particular symmetry, at any temperature $T \le T_d$ the ordering (in
free-energy) of the metastable states is the same they have (in
energy) at zero temperature \cite{KurchanParisiVirasoro}.  As a
consequence the metastable threshold states, those where the out of
equilibrium dynamics converges to, are the same at any temperature
below $T_d$.  However, for this very reason, the single-$p$-spin model
can be considered as pathological.  In more realistic systems, when
the temperature (or any other external parameter) varies, metastable
states may appear/disappear and their ordering is no longer preserved.
For such systems, many questions on the dynamical behaviour are still
open: for example, it is not clear what is the asymptotic state of the
off-equilibrium dynamics if the temperature varies during the
relaxation and whether such asymptotic state can be computed directly
from the thermodynamical measure (i.e.\ statically), without solving
the dynamical equations.  In this paper we try to give some answers to
the above points.

A last comment on the relevance of the present work concerns the
difference between a cooling and a quench.  The common experience
(exploited by the Simulated Annealing method) tells us that, in the
same amount of {\em finite} time, the cooling is able to reach lower
energies than the quench.  But what happens in the large times limit?
In the single-$p$-spin model the answer is simple: any cooling scheme,
included the quench, converges to threshold states, which are
$T$-independent.  On the contrary, if threshold states vary with
temperature, the answer was unclear and we show analytically that the
asymptotic state may depend on the cooling scheme.  The paper is
organized as follows: Section \ref{sec:model} describes the model we
want to study and summarizes what was already known about it; in
Section \ref{sec:dyn} we write and solve the equation for the out of
equilibrium dynamics, starting from a thermalized configuration; in
Section \ref{sec:statics} we report the results of the computation of
the number of {\sc tap} states, constrained to a fixed distance from a
reference state (more details are given in Appendix \ref{sec:app}) and
we compare these results with the solution of the dynamical equations;
finally in Section \ref{sec:conclusions} we summarize our results and
give some future perspective.

\section{The model}\label{sec:model}

We focus our attention on a modified version of the $p$-spin
Hamiltonian, the so-called multi-$p$-spin model, in which spins do
interact in $r$-uples with $r$ taking more than one value:
\begin{equation}
\mathcal{H}\{\underline\s\} = -\sum_r c_r \sum_{1 \le i_1 < \ldots <
  i_r \le N} J_{i_1 \ldots i_r} \s_{i_1} \ldots
  \s_{i_r}\;,\label{eq:ham}
\end{equation}
Using the overline for the disorder average, we have that
\begin{equation}
\overline{\mathcal{H}\{\underline\s\} \mathcal{H}\{\underline\t\}} =
\frac12 \sum_r c_r^2 q^r \equiv f(q)\;,
\end{equation}
where $q \equiv \sum_i \s_i \t_i / N$ is the overlap among
$\underline\s$ and $\underline\t$.  The single-$p$-spin corresponds to
$f(q)=q^p/2$.

The choice of the Hamiltonian (\ref{eq:ham}) is motivated by the
request of an exactly solvable dynamics, for which we need continuous
variables interacting in a fully-connected fashion.  Unlike the
single-$p$-spin case, in the multi-$p$-spin model there is level
crossing of metastable states by varying the temperature
\cite{RizzoYoshino}.
 
From a statical point of view, this model is characterized by the
presence of a large number of metastable states $\mathcal{N}(f)\sim
\exp[N \Sigma (f)]$.  The so-called {\it complexity} $\Sigma$ is an
increasing function of the free-energy $f$ which is zero at the lower
band edge $f_0$ and maximal for a certain value $f=f_{max}$.  For high
temperatures the Gibbs measure is dominated by the paramagnetic state
($m_i=0$), while for $T<T_d$ ($T_d$ being the dynamical critical
temperature), metastable states start to play a relevant role, much in
the same way as for the single-$p$-spin model.  In this region the
thermodynamic equilibrium is given by a class of metastable
`equilibrium' states with finite complexity (see e.g.\
Ref.~\onlinecite{FranzParisi}), the global free energy of the system
thus bearing a contribution from this state-related entropy, i.e.\ $-T
\ln Z = F = f_{eq} - T \Sigma(f_{eq})$.  Lowering still more the
temperature, the complexity of the equilibrium states decreases until
a point where it becomes zero and the lower band edge states, non
exponential in number, become dominant. The temperature where this
occurs, $T_s$, is the static transition temperature for this model, as
can be seen also by a direct computation of the partition function
with the replica method. The interpretation of this transition as an
`entropy crisis' for metastable states is particularly relevant when
comparing this model with real systems: indeed fragile glasses do
exhibit in this respect a very similar phenomenology.
 
The structure of metastable states can be investigated in much detail
by considering the {\sc tap} approach, where mean-field equations can
be formulated for the local magnetizations $m_i$ (at fixed disorder
realization), and stable solutions of these equations identified as
states of the system.  Recently, some novel intriguing features of
this formalism have emerged, according to which metastable states can
either satisfy a supersymmetry between fermionic and bosonic
integration variables \cite{SS}, either break it \cite{Aspelmeier}.
Supersymmetric (SS) states are very robust to external perturbations,
while supersymmetry breaking ones (SSB) are extremely fragile, and
even a small perturbation can dramatically change their number and
global structure \cite{cavity}.  Interestingly, the multi-$p$-spin
model addressed in this paper, contrary to the single-$p$ case,
exhibits states of both classes \cite{giulia} and allows a comparative
study of their role. In particular, states in the range $[f_0,f_{th}]$
are SS, while states with $f\in [f_{th},f_{max}] $ are SSB. The free
energy level $f_{th}$ separating the SS from the SSB region, that we
shall call {\it threshold energy}, also plays a relevant role in the
dynamical behaviour of this system.
 
Another important feature of metastable states, which is more relevant
for the questions we want to address, is their behaviour with changing
the temperature. For the single-$p$-spin spherical model, as
anticipated above, states can be transposed in temperature and their
energy ordering does not change. There is no birth/death of states
with varying the temperature, or, in other words, a {\sc tap} solution
at zero temperature persists when the temperature is turned on until
$T=T_d$.  In the multi-$p$-spin model this is not the case. To
investigate more explicitly this point, we need a method to `pin' out
a state and `follow' it with varying the temperature.  This can be
done by resorting to a dynamical analysis.

\section{The dynamics}\label{sec:dyn}

Given the Hamiltonian (\ref{eq:ham}) with a generic correlator $f(q)$,
it is possible to write the equations for the Langevin dynamics at
temperature $T=1/\b$ as
\begin{equation}
\frac{\partial \sigma_i(t)}{\partial t} = - \frac{\partial 
\mathcal{H}\{\underline\sigma\}}{\partial \sigma_i} +\eta(t)\;,
\label{dynamics}
\end{equation}
where $\eta(t)$ is a thermal Gaussian noise with zero mean and
variance
\begin{equation}
\quad \langle \eta(t) \eta(t')\rangle  = \frac{2}{\b}\delta(t-t')\ .
\end{equation}

Given the initial conditions this equation can be solved exactly using
the method of the generating functional \cite{DeDominicis}.
Self-consistent equations for the correlation function $C(t,t')=
\overline{\langle \sigma_i(t) \sigma_i(t') \rangle}$ and the response
function $R(t,t')=\overline{\partial \sigma_i(t)/\partial h_i(t')}$
read
\begin{eqnarray}\label{correlation}
\frac{ \partial C (t,t')}{\partial t}&=& -\mu(t) C(t,t') + \int_0^{t'}
\; ds f'[C(t,s)] R(t',s) + \nonumber \\
&+&\int_0^t \, ds R(t,s) \,f''[C(t,s)] C(s,t') + \b' f'[C(t,0)]
C(t',0) \\
\frac{\partial R(t,t')}{\partial t} &=& -\mu(t)R(t,t')\, + \int_0^{t'}
\; ds f''[C(t,s)] R(t,s) R(s,t')\label{response}
\end{eqnarray}
where $\mu(t)$ is a Lagrange multiplier implementing the spherical
constraint on the spins and obeys the dynamical equation
\begin{equation}
\mu(t) = \int_0^t \; ds f'[C(t,s)] R(t,s) + \int_0^t \, ds R(t,s)
\,f''[C(t,s)] C(s,t) +\frac{1}{\b}+ \b' f'[C(t,0)] C(t,0)
\end{equation}

The most studied case is the one where initial conditions are chosen
at random ($\b'=0$), and the system starts exploring the configuration
space from a high energy configuration. In this context, for example,
the first analytical complete treatment of aging behaviour has been
carried out for the single-$p$-spin \cite{CuKu}.

From our point of view, however, the most interesting situation is
another one.  If we choose the initial condition
$\underline\sigma(t=0)$ to belong to a given metastable state, then we
can let the system evolve and check whether the state is stable and
well-defined (in which case we expect an equilibrium-like relaxation
dynamics inside the state) or looses stability (exhibiting
off-equilibrium behaviour).  To this aim \cite{BarratFranzParisi}, we
may choose an initial condition thermalized at temperature $T'=1/\b'$,
i.e.
\begin{equation}
P\{\underline\sigma(0)\} = \frac{1}{Z} 
\exp{\left[ -\b'\mathcal{H}\{\underline\sigma(0)\}\right]}
\label{dynamics1}
\end{equation}
Indeed, since for $T_s < T' \le T_d$ the Boltzmann measure is
dominated by a class of metastable states, the distribution
(\ref{dynamics1}) naturally picks out a configuration
$\underline\sigma(0)$ which belongs to one of such states. Besides,
since the class of dominating states varies with the temperature, we
can use $T'$ to select the kind of state (i.e. energy, complexity and
self-overlap) we want the system to start in.

Summarizing, the dynamics that we are considering, described by
Eqs.~(\ref{dynamics}-\ref{dynamics1}), involves two distinct
temperatures.  The first one, $T$, controls the thermal noise and
therefore represents the temperature at which the dynamical evolution
takes place. The second one, $T'$ is used to force the system to start
into a given metastable state, and to select its properties.  We now
analyze the dynamical behaviour of the system by tuning these two
parameters.

\subsection{The quench}
The case $\b'=0$ corresponds to random initial conditions, that is to
a {\it quench}.  In this case the system undergoes a dynamical
transition at a critical temperature $T=T_d$ where the relaxation time
diverges (much in the same way as in the single-$p$-spin).  For
$T<T_d$ the dynamics exhibits aging and asymptotically reaches the
threshold states, which have energy density $E_{th}$ (corresponding to
free energy density $f_{th}$) and self-overlap $q_{th}=q_m$, where
$q_m(T)$ is the solution to the marginality condition
\begin{equation}
f''(q_m) (1-q_m)^2 = T^2\;.
\end{equation}
A similar behaviour occurs for any $T'>T_d$.

We note that the dynamics following a quench always converges to the
edge of the SS region, despite a larger number of SSB states are
present at higher energy densities ($E_{th}<E<E_{max}$).  At least two
explanations are possible: (i) SSB states are ``invisible'' for the
dynamics we have solved; (ii) SSB states are marginally unstable (they
have a finite number of zero modes in the thermodynamic limit) and
they are unable to trap the system during the relaxation
\cite{Aspelmeier}.

\subsection{The case $T_s < T' \le T_d$}
\begin{figure}
 \includegraphics[width=0.8\columnwidth]{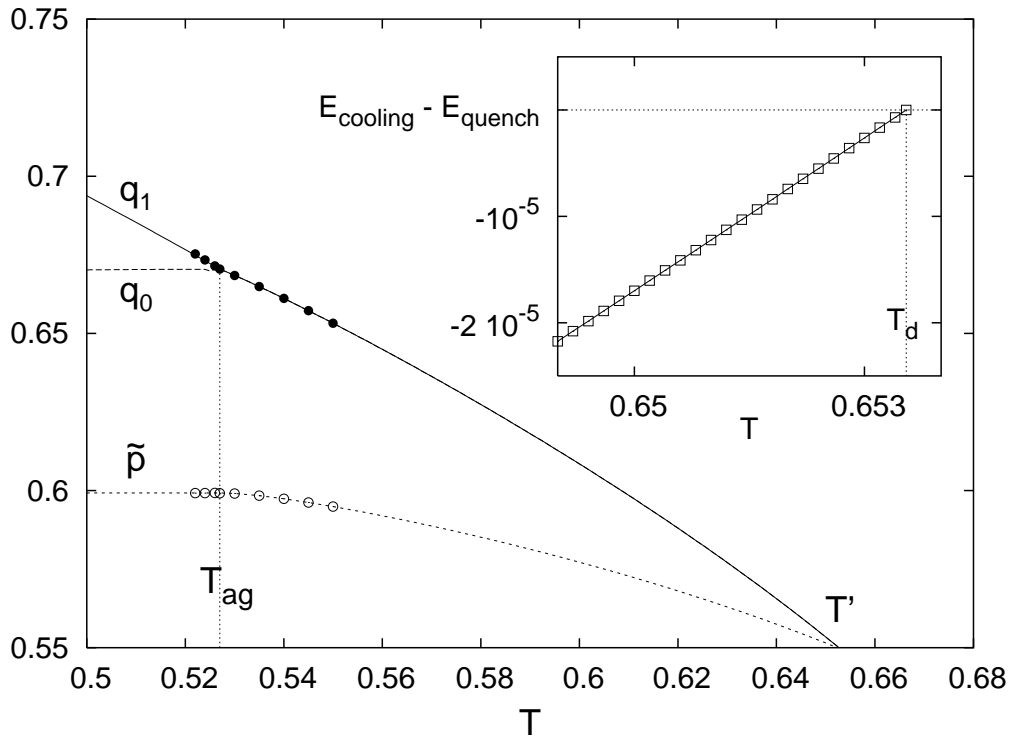}
 \caption{The dynamical parameters $q_1$, $q_0$ and $\pt$ (curves);
  the mutual overlap $q_{12}$ and the self-overlap obtained from the
  {\sc tap} complexity computation (points).  Here $f(q) = q^3/2 +
  (0.45)^2 q^4/2$.  Relevant temperature values are: $T_d=0.6543$,
  $T'=0.653$ and $\Trsb(0.653)=0.527$.  Inset: difference between
  asymptotic energies in a quench and a cooling (see text).}
 \label{fig:parameters}
\end{figure}

When $T_s < T' \le T_d$, the situation is more complex: as described above
the thermodynamic equilibrium is no longer given by the paramagnetic
state but rather by a set of metastable states with energy density $E
\in [E_0,E_{th}]$. Thus, the initial configuration belongs to one of
such states.  For $T=T'$ the system undergoes an equilibrium dynamics
in the state where it was at the starting time.  For $T<T'$ the
initial condition is out of equilibrium and, according to the value of
$T$, different dynamical behaviours can be observed. In particular, a
critical temperature $\Trsb(T')$ exists, such that:

i) For $\Trsb(T')\!< T < T'\!<\!T_d$ the system follows, at large
times, an equilibrium relaxation dynamics. Eqs.~(\ref{correlation})
and (\ref{response}) can be easily solved exploiting
time-translational invariance and the fluctuation-dissipation relation
between correlation and response.  The asymptotic regime is then fully
described by the two parameters
\begin{equation}
q_1 = \lim_{(t-t') \to \infty} \lim_{t'\to\infty} C(t,t') \quad
\mbox{and} \quad \pt = \lim_{t \rightarrow \infty} C(t,0)\;,
\label{equilibrium}
\end{equation}
which turn out to be different from zero \cite{BarratFranzParisi},
similarly to the single-$p$-spin model \cite{BarratBurioniMezard}.
The physical interpretation is clear: the system has been prepared
inside an equilibrium state at temperature $T'$; at temperature $T$
this state still exists, even if with slightly different features, and
the system dynamically relaxes into it. In this view, $q_1$ identifies
the self-overlap of the state at temperature $T$, while $\tilde p$
measures the overlap between the equilibrium state at $T'$, where the
initial configuration is placed, and the {\it same} state transposed
at temperature $T$.

ii) For $T<\Trsb(T')<T_d$ the dynamics remains out of equilibrium even
for large times, showing aging and violation of the time-translation
invariance.  Equations for the correlation and the response functions
can be written using the same scaling ansatz as the single-$p$-spin
model \cite{CuKu,BarratFranzParisi}.  For asymptotic but close times
[$t' \to \infty$, $t-t'=\mathcal{O}(1)$], time translation invariance
is recovered and the parameter $q_1$ can be defined as in
Eq.~(\ref{equilibrium}).  For asymptotic and well separated times [$t'
\to \infty$, $t/t'=\mathcal{O}(1)$], the correlation function scales
as $C(t,t') = \mathcal{C}(\lambda)$, with $\lambda \equiv t'/t \in
[0,1]$ and $\mathcal{C}(1) = q_1$ (the same scaling holding for the
response).  This regime defines another asymptotic parameter $q_0 =
\lim_{\lambda \to 0} \mathcal{C}(\lambda)$.  In this temperature
region, the asymptotic limit is fully described in terms of the three
parameters $q_1$, $q_0$ and $\pt$, together with the so-called {\it
fluctuation-dissipation ratio} $x_{dyn} \equiv T R(t,t') /
\partial_{t'}C(t,t')$ measuring the violation of the
fluctuation-dissipation relation.  The explicit equations for these
quantities read\footnote{Equations like (\ref{equazioni}) first
appeared in Ref.~\onlinecite{BarratFranzParisi}.  However, in that
paper there was a mistake in one of such equations and the
off-equilibrium dynamics was not solved.}
\begin{equation}
\left\{
\begin{array}{rcl}
1&=&\beta^2 f''(q_1)(1-q_1)^2\\
\frac{q_1}{\beta(1-q_1)}&=&\beta f'(q_1)(1-q_1)+\\
&&\beta x [q_1 f'(q_1)-q_0f'(q_0)] + \beta'\pt f'(\pt) \\
\frac{\pt}{\beta(1-q_1)}&=&\beta x \pt[f'(q_1)-f'(q_0)]+\beta' f'(\pt)\\
\frac{q_0}{\beta(1-q_1)}&=&\beta f'(q_0)(1-q_1)+\b x f'(q_0)(q_1-q_0)\\
&&+\beta x q_0 [f'(q_1)-f'(q_0)]+\beta'\pt f'(\pt)
\end{array}
\right.   
\label{equazioni}
\end{equation}
The parameters $q_0$, $q_1$ and $\pt$ are plotted in
Fig.~\ref{fig:parameters} for $f(q) = q^3/2 + (0.45)^2 q^4/2$ (the
same correlator used in Ref.~\onlinecite{giulia}).  $T'$ has been
chosen very close to $T_d$ in order to have a large $\Trsb$ value.

The first of equations (\ref{equazioni}) coincides with the
marginality condition obeyed by threshold states and defining
$q_m(T)$. However, the asymptotic dynamical energy
$E_{dyn}=\lim_{t\to\infty}E(t)$ that we obtain is different (and
lower) from the threshold energy $E_{th}$, indicating that the
asymptotic dynamics takes place in a marginal manifold below the
threshold states one.  This feature also holds when $T'=T_d$, which is
relevant for understanding the behaviour of a cooling in this system.
Indeed an infinitely slow cooling is able to reach thermal equilibrium
at any temperature above and at $T_d$ \footnote{Convergence to
equilibrium at $T_d$ is a subtle point: there are no $\mathcal{O}(N)$
barriers (states are marginal), and so the equilibration time is less
than exponential in $N$ (we expect it to grow polynomially with
$N$).}: thus for temperatures $T$ below $T_d$ an infinitely slow
cooling is roughly equivalent to a dynamics starting thermalized at
$T'=T_d$.  Our result then indicates that for this system the
asymptotic states reached with a cooling are {\em lower} than those
reached with a quench.  Actually the difference between these
asymptotic states is really very tiny (see inset of
Fig.~\ref{fig:parameters}).  To our knowledge this is the first
analytically solvable model showing up a dependence of the asymptotic
dynamical states on the cooling schedule.

Note also that the solution of Eqs.~(\ref{equazioni}) has $q_0\neq 0$.
This means that the system never decorrelates completely.  We are
observing aging {\em together} with a strong dynamic ergodicity
breaking, contrary to the weak-ergodicity breaking scenario analyzed
for this kind of models so far \cite{DynamicsReview}.  As long as $\pt
> 0$ the initial condition is not forgotten by the aging system and
thus the initial condition acts like a magnetic field, inducing $q_0 >
0$.

The change in the dynamical behavior at $T=\Trsb$ can be better
understood by noticing that $q_1 > q_m$ as long as $T>\Trsb$, and $q_1
= q_m$ for $T \le \Trsb$.  The simplest interpretation is that states
dominating the Gibbs measure at $T'$ are stable for $T>\Trsb(T')$, but
at $\Trsb(T')$ they become marginal, forming a manifold where the
system ages on.  For $T<\Trsb(T')$ these states become unstable and
the system keeps aging in a nearby critical manifold with $q_1=q_m$
and $q_0 < q_1$.

This interpretation also tells us that in the multi-$p$-spin model,
contrary to the single-$p$-spin, it is not possible to `follow' any
state from $T=0$ to finite temperature, or viceversa, because some
states loose stability as the temperature is varied and there may be
birth/death of states with temperature. Moreover, changing the
temperature the complexity $\Sigma$ varies along the dynamical
trajectory, implying that there is mixing of states.  The variation in
$\Sigma$ is very tiny: e.g.\ for $f(q) = q^3/2 + (0.45)^2 q^4/2$ is
$\mathcal{O}(10^{-5} \div 10^{-6})$ and for this reason it would be
hardly visible in a numerical simulation \cite{MontanariRicci}.

\section{The constrained complexity}\label{sec:statics}

To confirm the interpretation given above and to better understand the
dynamical behaviour, we can look more in details at the structure of
the metastable states in the region where the asymptotic dynamics
occurs. In particular, we can consider the following static
quantity. Given a reference state, that will be appropriately chosen
as the state to which the initial configuration of the dynamics
belongs, we compute the number of metastable states of given free
energy density that have fixed mutual overlap with it. Using the index
$1$ for the reference state and $2$ for the metastable states we wish
to count, we compute ${\mathcal N}(q_{12},f_2|f_1) \sim \exp[N
\Sigma(q_{12},f_2|f_1)]$, that is the number of states of free energy
density $f_2$ that have mutual overlap $q_{12}$ with a reference state
of free energy density $f_1$.  Temperatures are not written
explicitly, but it is assumed that $T_1=T'$ and $T_2=T$.

This computation can be performed in the {\sc tap} approach
\cite{TAP}, where metastable states are identified with local minima
of the mean-field energy functional $\FTAP\{\underline{m}\} =
H\{\underline{m}\} - 1/(2 \beta) \log(1-q) - \frac{\beta}{2} \left(
f(1) - f(q) -(1-q) f'(q) \right)$, where $q \equiv N^{-1} \sum_i
m_i^2$. In this case, the number of states reads
\begin{multline}
\mathcal{N}(q_{12},f_2|f_1) = \int \prod_i \Big( {\rm d}
m^{\scriptscriptstyle(1)}_i {\rm d} m^{\scriptscriptstyle(2)}_i
\delta(\partial_i \FTAP\{\underline{m}^{\scriptscriptstyle(1)}\})
\delta(\partial_i \FTAP\{\underline{m}^{\scriptscriptstyle(2)}\})\Big)
\left|{\rm det}\,\widehat{H}\{\underline{m}^{\scriptscriptstyle(1)}\}\right|\,
\left|{\rm det}\,\widehat{H}\{\underline{m}^{\scriptscriptstyle(2)}\}\right|\\
\delta(\FTAP\{\underline{m}^{\scriptscriptstyle(1)}\} - N f_1) \,
\delta(\FTAP\{\underline{m}^{\scriptscriptstyle(2)}\} - N f_2) \,
\delta(\underline{m}^{\scriptscriptstyle(1)}\cdot
\underline{m}^{\scriptscriptstyle(2)} - N q_{12}) \times
\mathcal{N}(f_1)^{-1}
\label{numerotap}
\end{multline}
In this expression the first two delta functions ensure that
$\underline{m}^{\scriptscriptstyle(1)}$ and
$\underline{m}^{\scriptscriptstyle(2)}$ are stationary points of
$\FTAP$, ($\widehat{H}_{ij}\{\underline m\} \equiv
\partial_i\partial_j F_\text{\sc tap}\{\underline m\}$ being the
Hessian in the appropriate normalization factor), and imply that we
are looking at solutions of the mean-field equations (i.e.\ states).
The other delta functions fix respectively the free energy densities
and the mutual overlap of the solutions we wish to count.  Note also
that we have divided by the number of metastable states with free
energy $f_1$ in order to get the number of states of free energy $f_2$
with overlap $q_{12}$ with a {\it given} reference state of kind $1$
(otherwise we would have gotten the number of pairs).

The entropy related to (\ref{numerotap}), also called {\it constrained
complexity}, can be computed using standard techniques
\cite{CavagnaGiardinaParisi,Barbara} within the annealed
approximation, which is in general adequate to treat large free
energies $f_1, f_2 \sim f_{th}$ (see Appendix \ref{sec:app}).
Alternatively, the entropy can be computed as the Legendre transform
of a constrained thermodynamic free energy
\cite{Monasson,BarratFranzParisi}. The two results coincide within
numerical precision.

\begin{figure}
  \includegraphics[width=0.8\columnwidth]{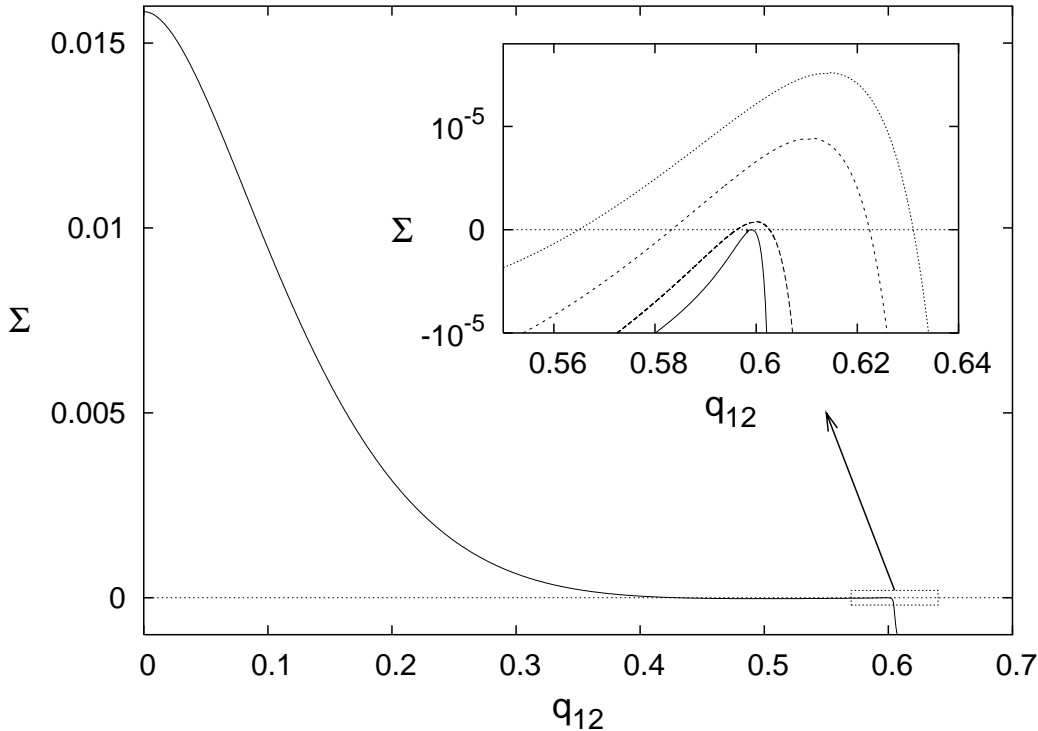}
  \caption{Constrained complexity as a function of the mutual overlap
    for $T\!= 0.53$ and $T'\!= 0.653$ ($\Trsb\!= 0.527$). Inset: the
    behaviour of the secondary peak for different temperatures, from
    bottom to top, $T\!= 0.53$, $T\!= 0.5$, $T\!= 0.4$, $T\!= 0.35$.}
  \label{fig:complexity}
\end{figure}

Let us now use this constrained entropy to investigate the structure
of the phase space sampled by the asymptotic dynamics. To this end,
let us fix $f_1=f_{eq}(\beta')$ and $f_2=f_{dyn}(\beta)\equiv
E_{dyn}(\beta)-T S(E_{dyn})$. That is, the reference state is chosen
as the state where the dynamics has been started in (an equilibrium
state at temperature $T'$), while the states to be counted have energy
density equal to the asymptotic dynamical energy.  The behaviour of
$\Sigma$ as a function of $q_{12}$ is displayed in
Fig.~\ref{fig:complexity}.  We see that two different situations occur
above and below $\Trsb$.

For $T>\Trsb(T')$ the constrained complexity is positive and
decreasing with increasing $q_{12}$, (as discussed for the
single-$p$-spin in Ref.~\onlinecite{CavagnaGiardinaParisi}), it
becomes negative at some value of the mutual overlap and touches back
the zero axis for $q_{12}=\pt$ (see the lowest curve in the inset of
Fig.~\ref{fig:complexity}), with $\pt$ given by the dynamical
equations~(\ref{equazioni}).  The interpretation is
straightforward. For small overlaps we are counting states in a very
large manifold, and we thus find many of them. As $q_{12}$ decreases,
this manifold becomes smaller and, consequently, the number of counted
states decreases until when it becomes zero (negative
complexity). However, if we still increase the overlap, looking closer
to the reference state, at some point we are bound to find the state
itself. This is signaled by the zero value of the complexity at
$q_{12} = \pt$, which therefore represents the overlap between the
reference state and the {\it same} state evolved at temperature $T$,
consistently with the dynamical interpretation.  In this point one
also has $q^{\scriptscriptstyle(2)} \equiv
\underline{m}^{\scriptscriptstyle(2)} \cdot
\underline{m}^{\scriptscriptstyle(2)} / N = q_1$, with $q_1$ given
again by the dynamical equations.  Please note that
$q^{\scriptscriptstyle(2)}=q_1$ is not the typical value for {\sc tap}
states at temperature $T$ and free-energy $f_{dyn}$; so the dynamics
is restricted to a set of sub-dominant states, that can be selected by
constraining the {\sc tap} measure as in (\ref{numerotap}).  The
interpretation is straightforward: the $T'$ equilibrium state has
evolved in a slightly modified state at temperature $T$, which has
overlap $\pt$ with the original one.  This is the {\it only} state
(i.e.\ $\Sigma=0$) that we count at temperature $T$ when fixing
$q_{12}=\pt$.  Note that this state is stable (by computing the
replicon) and it is ``isolated'', that is the $\Sigma$ curve is
negative in the $(q_{12},f_2)$ plane around the point $(\pt,f_{dyn})$.

\begin{figure}
  \includegraphics[width=0.8\columnwidth]{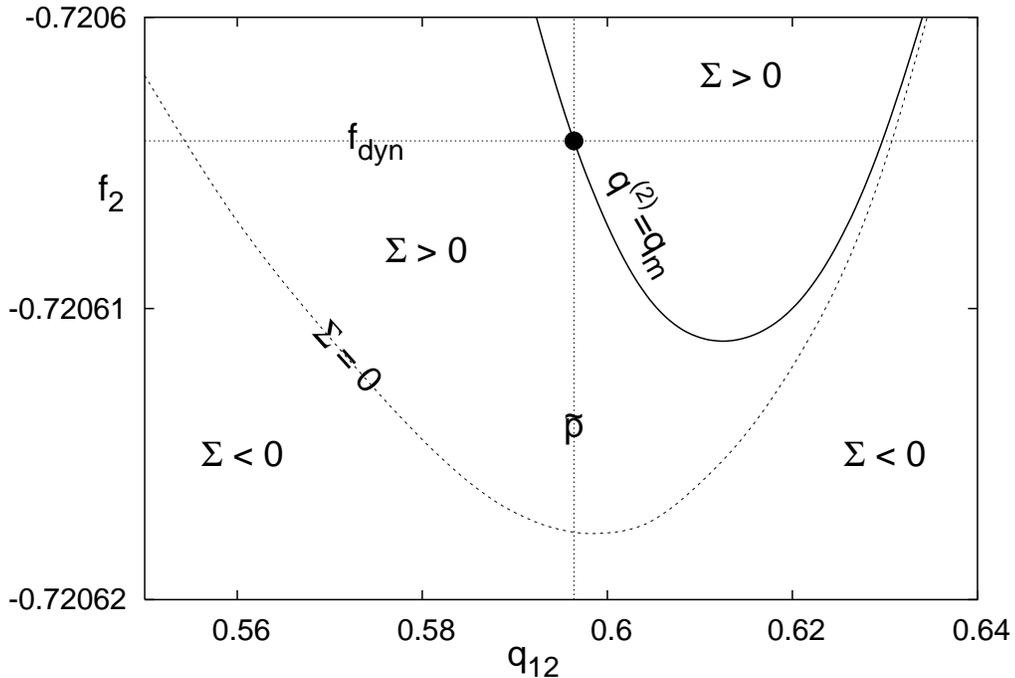}
  \caption{Constrained complexity in the $(q_{12},f_2)$ plane, with
    $f_1=f_{eq}(T'=0.653)$ and $T=0.35$.}
  \label{fig:region}
\end{figure}

For $T<\Trsb(T')$ the secondary peak of the constrained complexity
becomes positive (see inset of Fig.~\ref{fig:complexity}) and the
$\Sigma>0$ region in the $(q_{12},f_2)$ plane (with $f_1$ fixed) is
shown in Fig.~\ref{fig:region}.  The $T'$ equilibrium state opens up at
$T=\Trsb(T')$ and a non trivial structure of metastable states appears
close to where the dynamics is taking place; these states are
responsible for the aging behavior, but it is still unclear which are
the thermodynamical parameters of the dynamical asymptotic
states. This is the main question when trying to describe the
dynamical behavior in terms of static observations.

We know that for $T<\Trsb$ the dynamics is taking place on a marginal
manifold, so we can fix $q^{\scriptscriptstyle(2)}=q_m(T)$: states
with this self-overlap are found along the full line in
Fig.\ref{fig:region}. All the points along this line are possible
candidates for the asymptote of the dynamics, but understanding which
one is actually chosen during system relaxation is a difficult task.

Consistently with the dynamical computation the point $(\pt,f_{dyn})$
is always on the line. Moreover, at this point, $x_{st}=x_{dyn}$
holds, where $x_{st}(q_{12},f_2) \equiv T \partial_f
\Sigma(q_{12},f,f_1)|_{f=f_2}$ and $x_{dyn}$ is the dynamical
fluctuation-dissipation ratio.

It seems that at least one dynamically computed quantity must be
plugged in the static computation to predict the asymptotic states:
this can be equivalently $q_{12}=\pt$ or $x_{st}=x_{dyn}$ (for a
quench the computation is easier: starting with a random configuration
one has $q_{12}=0$ by definition).  It would be very useful to find an
extremizing principle to select, among all the candidate {\sc tap}
states, those which are actually reached by the constrained out of
equilibrium dynamics.

\section{Summary and perspectives}\label{sec:conclusions}

The spherical multi-$p$-spin model defined by the Hamiltonian
(\ref{eq:ham}) has the nice properties of being exactly solvable
(thanks to its continuous variables), while showing non-trivial
dependence on temperature of its states (birth, death and level
crossing).  These features makes the model more realistic than other
mean-field models and a perfect candidate for studying glassy
relaxation under variations of temperature.

We have performed such a study finding several interesting analytical
results.  (i) The relaxation at any temperature converges to \tap\
states satisfying the supersymmetry between fermionic and bosonic
integration variables. In order to understand whether \tap\ states
breaking the supersymmetry are relevant for {\em finite times}
dynamics, the method described in Ref.~\onlinecite{BiroliKurchan}
could be applied to the present model.  (ii) Energies reached by a
cooling are lower than those reached by a quench.  This result is
based on the assumption than an infinitely slow cooling equilibrates
at any temperature $T \ge T_d$, which needs to be improved.  (iii) The
solution to the dynamical equations is consistent with the constrained
complexity of \tap\ states computed thermodynamically: starting from a
thermalized configuration and lowering the temperature, the system
starts aging when the state it belongs to becomes marginally
unstable. For lower temperatures, states where aging is taking place
cannot be predicted solely from the constrained complexity; a new
extremizing principle is needed in order to make the connection
between static and dynamic computations.

\acknowledgments We thank A. Annibale, A. Cavagna and G. Gualdi for
useful discussions.  This work has been supported by the EEC's FP6 IST
Programme under contract IST-001935, EVERGROW, and by the ECC's HPP
under contracts HPRN-CT-2002-00307 (DYGLAGEMEM) and HPRN-CT-2002-00319
(STIPCO).

\appendix
\section{The constrained complexity via the {\sc tap} approach}
\label{sec:app}

Starting from Eq.~(\ref{numerotap}), the computation of the
constrained complexity can be performed with standard techniques.  The
annealed computation is the simpler one, since it involves averaging
directly the number of solutions over the quenched disorder, rather
than its logarithm. Replicas are therefore not needed.

To proceed, we introduce bosonic representations for the delta
functions appearing in Eq.~(\ref{numerotap}), and fermionic
representations for the two determinants (the modulus can be safely
disregarded for this model, since one can show that minima dominate
over saddles in the relevant free energy density range
\cite{saddles}). In this way, we get
\begin{equation}
  \mathcal{N}(q_{12},f_2,\b|f_1,\b')  = \frac{
\int D[\underline{m}^{\scriptscriptstyle(1)},
  \underline{m}^{\scriptscriptstyle(2)}, 
  \underline{x}^{\scriptscriptstyle(1)}, 
  \underline{x}^{\scriptscriptstyle(2)},
  \underline{\bar\psi}^{\scriptscriptstyle(1)},
  \underline{\psi}^{\scriptscriptstyle(1)}, 
  \underline{\bar\psi}^{\scriptscriptstyle(2)},
  \underline{\psi}^{\scriptscriptstyle(2)},
  u_1 , u_2, w]\;e^{S_{tot}}}{\mathcal{N}(f_1,\b')}
\label{eq:A1}
\end{equation} 
with
\begin{equation}
S_{tot}  = 
\beta' S(\underline{m}^{\scriptscriptstyle(1)},
\underline{x}^{\scriptscriptstyle(1)},
\underline{\bar\psi}^{\scriptscriptstyle(1)},
\underline{\psi}^{\scriptscriptstyle(1)}, u_1; f_1, \b')
+ \beta S(\underline{m}^{\scriptscriptstyle(2)},
\underline{x}^{\scriptscriptstyle(2)},
\underline{\bar\psi}^{\scriptscriptstyle(2)},
\underline{\psi}^{\scriptscriptstyle(2)}), u_2; f_2, \b)
+ w (N q_{12} - m^{\scriptscriptstyle(1)} \cdot
m^{\scriptscriptstyle(2)}) \;,
\end{equation}
\begin{equation}
S(\underline{m}, \underline{x},  \underline{\bar\psi}, 
\underline{\psi}, u; f, \b) = \sum_i x_i \partial_i
\FTAP(\underline{m}, \b) + \sum_{ij} \bar\psi_i
\partial_i\partial_j \FTAP(\underline{m}, \b) \psi_j +
u (\FTAP(\underline{m},\b) - N f)\;,
\end{equation}
and
\begin{equation}
\FTAP(\underline{m},\b) = H\{\underline{m}\} - 1/(2 \beta) \log(1-q) -
\frac{\beta}{2} \left[ f(1) - f(q) -(1-q) f'(q) \right]\;,
\end{equation}
where $\underline{x}^{(1,2)}$ are the Lagrange multipliers enforcing
the {\sc tap} equations, $u_{1,2}$ are those enforcing the free energy
constraint, $w$ is the one for the mutual overlap constraint, and
$\underline\psi^{(1,2)},\;\underline{\bar\psi}^{(1,2)}$ are the
Grassman variables used to represent the determinants.  As usual,
$\partial_i$ is the shorthand notation for $\partial/\partial_{m_i}$.

After averaging over the disorder the numerator and the denominator of
expression (\ref{eq:A1}), consistently with the annealed
approximation, site dependent variables can be integrated out, leaving
an effective action which only depends on global variables:
\begin{equation}
\overline{\mathcal{N}}=\int D[\Omega,\omega,u_1,u_2,w] \exp\left[ N
S_{eff}(\Omega,\omega,u_1,u_2,w; q_{12},f_2,\b,f_1,\b') - N
\Sigma(f_1,\b') \right]\;,
\end{equation}
where $\Omega = [q_1,q_2, B_1,B_2,R_1,R_2, B_{12}, B_{21}]$, defined
by ($a,b=1,2$)
\begin{eqnarray}
q_a &=& (\underline{m}^{\scriptscriptstyle(a)} \cdot
\underline{m}^{\scriptscriptstyle(a)})/N \nonumber \\
B_a &=& (\underline{m}^{\scriptscriptstyle(a)} \cdot
\underline{x}^{\scriptscriptstyle(a)})/N \nonumber \\
R_a &=&(\underline{\bar\psi}^{\scriptscriptstyle(a)} \cdot
\underline{\psi}^{\scriptscriptstyle(a)})/N \nonumber \\
B_{ab} &=& (\underline{m}^{\scriptscriptstyle(a)}\cdot
\underline{x}^{\scriptscriptstyle(b)})/N
\end{eqnarray}
and $\omega = [\lambda_1,\lambda_2,b_1,b_2,r_1,r_2,b_{12},b_{21}]$
are the corresponding Lagrange multipliers.

The explitic expression for the effective action is the following:
\begin{equation}
\begin{split}
S_{{eff}} & = \frac{1}{2} \log z - \frac{1}{2} \log \Xi +
\log (2 \beta' g'(q_1) + r_1) + \log (2\beta g'(q_2) + r_2)\\
& + \beta' u_1(g(q_1) - f_1) + \beta u_2 (g(q_2) - f_2) +
\frac{{\beta'}^2}{2}(B_1^2 - R_1^2)f''(q_1) +
\frac{\beta^2}{2} (B_2^2 - R_2^2) f''(q_2)\\
& + \frac{{\beta'}^2}{2} u_1^2 f(q_1) + \frac{\beta^2}{2} u_2^2
f(q_2) + \beta \beta' B_{12} B_{21} f''(q_{12}) +
\beta \beta' u_1 u_2 f(q_{12})\\
& - r_1 R_1 - r_2 R_2 - b_1 B_1 - b_2 B_2 - b_{12} B_{12} - b_{21}
B_{21} - \lambda_1 q_1 - \lambda_2 q_2 - w q_{12}\; ,
\label{Seff}
\end{split}
\end{equation}
with
\begin{align*}
z & = (\beta\beta')^2 (f'(q_1) f'(q_2) - (f'(q_{12}))^2)\\
\Xi & = (l_{11} - 2z \lambda_1)(l_{22} - 2z \lambda_2) - (l_{12} - zw)^2\\
g(q) & =  - 1/(2 \beta) \log(1-q) - \frac{\beta}{2} \left[
f(1) - f(q) -(1-q) f'(q) \right]\;,
\end{align*}
\begin{align*}
l_{11} & = \beta^2 f'(q_2) d_{11}^2 - 2 \beta \beta'
f'(q_{12})d_{11} d_{21} + {\beta'}^2 f'(q_1) d_{21}^2,\\
l_{12} & = \beta^2 f'(q_2) d_{11} d_{12} - \beta \beta'
f'(q_{12})(d_{11} d_{22} + d_{21} d_{12}) + {\beta'}^2 f'(q_1)
d_{21} d_{22},\\
l_{22} & = \beta^2 f'(q_2) d_{12}^2 - 2 \beta \beta' f' (q_{12})
d_{12} d_{22} + {\beta'}^2 f'(q_1) d_{22}^2,
\end{align*}
\begin{align*}
d_{11} & = 2 \beta' A(q_1) + b_1 + {\beta'}^2 u_1 f'(q_1),\\
d_{12} & = b_{12} + \beta \beta' u_2 f'(q_{12}),\\
d_{21} & = b_{21} + \beta \beta' u_1 f'(q_{12}),\\
d_{22} & = 2 \beta A(q_2) + b_2 + \beta^2 u_2 f'(q_2).
\end{align*}

The effective action (\ref{Seff}) has to be extremized with respect to
all the integration variables in order to obtain the constrained
complexity as $\Sigma(q_{12},f_2,\b,f_1,\b') = S^{\,{\rm
extr}}_{eff}(q_{12},f_2,\b,f_1,\b') - \Sigma(f_1, \beta')$.

\end{document}